\documentclass[letterpaper,12pt]{article}

\usepackage{graphicx}
\usepackage{amsmath}
\newcommand{\be}{\begin{equation}}
\newcommand{\ee}{\end{equation}}
\newcommand{\bea}{\begin{eqnarray}}
\newcommand{\eea}{\end{eqnarray}}
\newcommand{\beaa}{\begin{eqnarray*}}
\newcommand{\eeaa}{\end{eqnarray*}}

\newcommand{\BB}{{{\rm I} \kern -2pt \rlap {\rm B} \kern +8pt}}

\advance\textheight by \topskip
\makeatletter

\@addtoreset{equation}{section}

\def\section{\@startsection {section}{1}{\z@}{-3.5ex plus -1ex minus
 -.2ex}{2.3ex plus .2ex}{\large\bf\centering}}
\def\subsection{\@startsection{subsection}{2}{\z@}{-3.25ex plus%
 -1ex minus -.2ex}{1.5ex plus .2ex}{\bf}}
\def\subsubsection{\@startsection{subsubsection}{3}{\z@}{-3.25ex plus%
 -1ex minus -.2ex}{1.5ex plus .2ex}{\sl}}
\makeatother

\begin{document}

\begin{titlepage}

\begin{center}
{\Large {\bf On noncommutative sinh-Gordon equation}}
\\\vspace{1.5in} {\large U. Saleem\footnote{%
usman\_physics@yahoo.com}, M. Siddiq\footnote{%
mohsin\_pu@yahoo.com}\footnote{%
On study leave from PRD (PINSTECH) Islamabad, Pakistan} and M. Hassan\footnote{%
mhassan@physics.pu.edu.pk} }\vspace{0.15in}

{\small{\it Department of Physics,\\ University of the Punjab,\\
Quaid-e-Azam Campus,\\ Lahore-54590, Pakistan.}}

\end{center}

\vspace{1cm}

\begin{abstract}
We give a noncommutative extension of sinh-Gordon equation. We
generalize a linear system and Lax representation of the
sinh-Gordon equation in noncommutative space. This generalization
gives a noncommutative version of the sinh-Gordon equation with
extra constraints, which can be expressed as global conserved
currents.
\end{abstract}
\vspace{1cm} PACS: 11.10.Nx, 02.30.Ik
\end{titlepage}

\section{Introduction}

Noncommutative geometry has widely been used in the study of
integrable field theories (IFTs) since the last decade
\cite{conne}-\cite{Muller4}. The noncommutative version of
integrable field theories (nc-IFTs) is obtained by replacing
ordinary product with $\star - $ star product. In the commutative
limit, these noncommutative theories reduce to ordinary field
theories. Noncommutative version of different integrable models,
such as principal chiral model with and without a Wess-Zumino
term, sine-Gordon and sinh-Gordon equations, Korteweg de Vries
(KdV) equation, Boussinesq equation, Kadomtsev-Petviashvili (KP)
equation, Sawada-Kotera equation, nonlinear
Schrodinger equation and Burgers equation, have been studied \cite{conne}-%
\cite{Muller4}.

The noncommutativity of space is characterized by
\[
\left[ x^{i},x^{j}\right] =\mbox{i} \theta ^{ij},
\]
where $\theta ^{ij}$ is a constant second rank tensor, called
parameter of noncommutativity. The $\star -$ star product of two
functions in noncommutative spaces is given by
\[
\left( f\star g\right) (x)=f(x)g(x)+\frac{\mbox{i}\theta
^{ij}}{2}\partial _{i}f(x)\partial _{j}g(x)+\vartheta (\theta
^{2}),\qquad \mbox{where $\partial _{i}=\frac{\partial }{\partial
x^{i}}.$}
\]
The $\star -$ star product obeys following properties:
\begin{eqnarray*}
f\star I &=&f=I\star f, \\
f\star \left( g\star h\right)  &=&\left( f\star g\right) \star h.
\end{eqnarray*}
The integration of two functions is \cite{Marco1}
\[
\int d^{D}xf(x)\star g(x)=\int d^{D}xf(x)g(x),
\]
where the integration is taken in all noncommutative directions.

The noncommutativity in $(1+1) -$ dimensional space-time is
defined as \cite{Hamanaka1}-\cite{Hamanaka3}
\[
[t,x]=\mbox{i}\theta.
\]
The $\star -$ star product of two functions in $(1+1)$-dimensional
noncommutative space is given by \cite{Hamanaka1}-\cite{Hamanaka3}
\[
\left( f\star g\right) (t,x)=f(t,x)g(t,x)+\frac{\mbox{i}\theta
}{2}\left. \left(
\partial _{t^{\prime }}\partial _{x^{\prime \prime }}-\partial _{t^{\prime
\prime }}\partial _{x^{\prime }}\right) f(t^{\prime },x^{\prime
})g(t^{\prime \prime },x^{\prime \prime })\right|_{t=t^{\prime
}=t^{\prime \prime };\,\ x=x^{\prime }=x^{\prime \prime }}
+\vartheta (\theta ^{2}),
\]
with $\partial _{t}=\frac{\partial }{\partial t}$. In what
follows, we present a noncommutative version of sinh-Gordon
equation and investigate the noncommutative version of its
zero-curvature and Lax representations. The integrability
condition of the linear system and the Lax equation gives rise to
a noncommutative sinh-Gordon equation with some extra constraints.

In section \ref{lin}, we give noncommutative generalization of a
linear system whose compatibility condition is the noncommutative
sinh-Gordon equation. In section \ref{laxpair}, we present a
noncommutative version of Lax representation of the noncommutative
sinh-Gordon equation. In Section \ref{per}, we expand the fields
perturbatively and obtain zeroth and first order sinh-Gordon
equations, the associated linear system and a set of parametric
B\"{a}cklund transformation (BT) of the sinh-Gordon equation. It
has been shown that the compatibility condition of the associated
linear system and the B\"{a}cklund transformation (BT) is the
sinh-Gordon equation at the perturbative level. Section \ref{con},
contains our conclusions.

\section{Linear System for Noncommutative sinh-Gordon
Equation}\label{lin}

In this section we discuss the integrability of noncommutative
extension of the sinh-Gordon equation. We start with an associated
linear system of the equation different from the one given in
Ref.\cite{Marco1} and show that its compatibility condition is the
noncommutative sinh-Gordon equation along with some constraints.
The constraints obtained here are different from those obtained in
Ref.\cite{Marco1}. These constraints are also shown to be
expressed as conserved global currents.

In general, a nonlinear evolution equation solvable by inverse
scattering method can be expressed as a compatibility condition of
a set of linear differential equations. The associated linear
system can be related to the isospectral problem and the Lax
representation. We now write a linear system whose compatibility
condition gives the noncommutative version of the sinh-Gordon
equation. The linear system of the sinh-Gordon equation is
\begin{equation}
\partial _{\pm }u=A_{\pm }^{\star }\star u,  \label{linear1}
\end{equation}
where $A_{\pm }^{\star }$ are
\begin{eqnarray*}
A_{+}^{\star } &=&\left(
\begin{array}{ll}
\,-\mbox{i}\lambda  & \frac{\beta}{2}\partial _{+}\varphi  \\
\frac{\beta}{2}\partial _{+}\varphi  & \mbox{i}\lambda
\end{array}
\right) , \\
A_{-}^{\star } &=&\frac{\mbox{i}m^2}{4\lambda }\left(
\begin{array}{ll}
\cosh _{\star }\beta\varphi  & -\sinh _{\star }\beta\varphi  \\
\sinh _{\star }\beta\varphi  & -\cosh _{\star }\beta\varphi
\end{array}
\right) .
\end{eqnarray*}
with $\varphi$ a real valued function and $\beta$, $m$ are some
positive parameters. The compatibility condition of the linear
system (\ref{linear1}) in noncommutative space is the
zero-curvature condition:
\[
\left[ \partial _{+}-A_{+}^{\star },\partial _{-}-A_{-}^{\star
}\right] _{\star }\equiv \partial _{-}A_{+}^{\star }-\partial
_{+}A_{-}^{\star }+\left[ A_{+}^{\star },A_{-}^{\star }\right]
_{\star }=0,
\]
where $\left[ A_{+}^{\star },A_{-}^{\star }\right]= A_{+}^{\star
}\star A_{-}^{\star }- A_{-}^{\star }\star A_{+}^{\star }$ is a
commutator in noncommutative space. The above compatibility
condition gives rise to the noncommutative sinh-Gordon equation
and some extra constraints
\begin{eqnarray*}
&&\left. \partial _{-}\partial _{+}\varphi =\frac{m^2}{\beta
}\sinh _{\star }\beta\varphi ,\right.
\\
&&\left. \partial _{+}\left( \cosh _{\star }\beta\varphi \right) -\frac{\beta}{2}%
(\sinh _{\star }\beta\varphi \star \partial _{+}\varphi +\sinh
_{\star }\beta\varphi
\star \partial _{+}\varphi )=0,\right.  \\
&&\left. \partial _{+}\left( \sinh _{\star }\beta\varphi \right) -\frac{\beta}{2}%
(\cosh _{\star }\beta\varphi \star \partial _{+}\varphi +\cosh
_{\star }\beta\varphi \star \partial _{+}\varphi )=0.\right.
\end{eqnarray*}
These constraints become total derivatives or as global conserved
currents. We note that in the limit $\theta \rightarrow 0$, the
first equation becomes the ordinary sinh-Gordon equation and extra
constraints vanish, the $\star-$ product becomes the usual
product.

\section{Lax Representation for Noncommutative sinh-Gordon
Equation}\label{laxpair}

In this section we present the Lax representation of the
noncommutative sinh-Gordon equation and find the corresponding Lax
equations.

The eigenvalue equation for a Lax operator $L_{\pm}$ is
\[
L_{\pm }\Psi =\lambda \Psi ,
\]
where $L_{\pm }$ is given by
\[
L_{\pm }=\left(
\begin{array}{ll}
\,\,\,-\mbox{i}\partial _{\pm } & \frac{\beta}{2}\partial _{\pm}\varphi  \\
\frac{\beta}{2}\partial _{\pm}\varphi  &
\,\,\,\,\,\,\mbox{i}\partial _{\pm }
\end{array}
\right) ,
\]
and
\[
\Psi =\left(
\begin{array}{l}
\Psi _{1} \\
\Psi _{2}
\end{array}
\right).
\]
The corresponding Lax equations are
\begin{equation}
\partial _{\mp }L_{\pm }=\left[ L_{\pm },M\right] _{\star },  \label{Lax}
\end{equation}
with
\[
M=\frac{\mbox{i}m^2}{4\lambda }\left(
\begin{array}{ll}
\cosh _{\star }\beta\varphi  & -\sinh _{\star }\beta\varphi  \\
\sinh _{\star }\beta\varphi  & -\cosh _{\star }\beta\varphi
\end{array}
\right) .
\]
The Lax equation gives noncommutative sinh-Gordon equation along
with some constraints
\begin{eqnarray*}
&&\left. \partial _{-}\partial _{+}\varphi =\frac{m^2}{\beta}\sinh
_{\star }\beta\varphi ,\right.
\\
&&\left. \partial _{\pm }\left( \cosh _{\star }\beta\varphi \right)-\frac{\beta}{2}%
(\sinh _{\star }\beta\varphi \star \partial _{\pm }\varphi +\sinh
_{\star
}\beta\varphi \star \partial _{\pm }\varphi )=0,\right.  \\
&&\left. \partial _{\pm }\left( \sinh _{\star }\beta\varphi \right) -\frac{\beta}{2}%
(\cosh _{\star }\beta\varphi \star \partial _{\pm }\varphi +\cosh
_{\star }\beta\varphi \star \partial _{\pm }\varphi )=0.\right.
\end{eqnarray*}
These constraints become total derivatives (global currents). The
first equation in the limit when the noncommutativity parameter
reduces to zero becomes an ordinary sinh-Gordon equation and the
constraints of the model vanish.

\section{Perturbative Expansion of sinh-Gordon Model}\label{per}

We can expand the field in power series of noncommutative
parameter $\theta$. The expansion of the field $\varphi $ up to
first order is
\[
\varphi =\varphi ^{(0)}+\theta \varphi ^{(1)}.
\]
With this expansion we can obtain two pairs of the sinh-Gordon
equation
\begin{eqnarray}
\partial _{+}\partial _{-}\varphi ^{(0)} &=&\frac{m^2}{\beta}\sinh \beta\varphi ^{(0)},
\label{sine1} \\
\partial _{+}\partial _{-}\varphi ^{(1)} &=&m^2\varphi ^{(1)}\cosh \beta\varphi
^{(0)}.  \label{sine2}
\end{eqnarray}
The linear system for equation (\ref{sine1}) becomes
\begin{equation}
\partial _{\pm }u^{(0)}=A_{\pm }^{(0)}u^{(0)},  \label{Plinear1}
\end{equation}
where $A_{\pm }^{(0)}$ are given by
\begin{eqnarray*}
A_{+}^{(0)} &=&\left(
\begin{array}{ll}
-\mbox{i}\lambda  & \frac{\beta}{2}\partial _{+}\varphi ^{(0)} \\
\frac{\beta}{2}\partial _{+}\varphi ^{(0)} &
\,\,\,\,\,\,\,\,\,\,\mbox{i}\lambda
\end{array}
\right) , \\
A_{-}^{(0)} &=&\frac{\mbox{i}m^2}{4\lambda }\left(
\begin{array}{ll}
\cosh \beta\varphi ^{(0)} & -\sinh \beta\varphi
^{(0)} \\
\sinh \beta\varphi ^{(0)} & -\cosh \beta\varphi ^{(0)}
\end{array}
\right) .
\end{eqnarray*}
The compatibility condition for the linear system (\ref{Plinear1})
is the zeroth order zero-curvature condition:
\[
\left[ \partial _{+}-A_{+}^{(0)},\partial _{-}-A_{-}^{(0)}\right]
\equiv
\partial _{-}A_{+}^{(0)}-\partial _{+}A_{-}^{(0)}+\left[
A_{+}^{(0)},A_{-}^{(0)}\right] =0.
\]
The linear system of the equation for first order reads
\begin{eqnarray}
\partial _{+}u^{(1)} &=&A_{+}^{(0)}u^{(1)}+A_{+}^{(1)}u^{(0)},
\label{SLinea} \\
\partial _{-}u^{(1)} &=&A_{-}^{(0)}u^{(1)}+A_{-}^{(1)}u^{(0)},  \nonumber
\end{eqnarray}
where $A_{\pm }^{(1)}$ are given by
\begin{eqnarray*}
A_{+}^{(1)} &=&\left(
\begin{array}{ll}
0 & \frac{\beta}{2}\partial _{+}\varphi ^{(1)} \\
\frac{\beta}{2}\partial _{+}\varphi ^{(1)} & \,\,\,\,\,\,\,\,\,\,0
\end{array}
\right) , \\
A_{-}^{(1)} &=&\frac{\mbox{i}m^2}{4\lambda }\left(
\begin{array}{ll}
\beta\varphi ^{(1)}\sinh \beta\varphi ^{(0)} & -%
\beta\varphi ^{(1)}\cosh \beta\varphi ^{(0)} \\
\beta\varphi ^{(1)}\cosh \beta \varphi ^{(0)} & -%
\beta\varphi ^{(1)}\sinh \beta\varphi ^{(0)}
\end{array}
\right) .
\end{eqnarray*}
The compatibility condition for the linear system (\ref{SLinea})
is
\[
\left( \partial _{-}A_{+}^{(0)}-\partial _{+}A_{-}^{(0)}+\left[
A_{+}^{(0)},A_{-}^{(0)}\right] \right) u^{(1)}+\left( \partial
_{-}A_{+}^{(1)}-\partial _{+}A_{-}^{(1)}+\left[
A_{+}^{(1)},A_{-}^{(0)}\right] +\left[
A_{+}^{(0)},A_{-}^{(1)}\right] \right) u^{(0)}=0.
\]
The B\"{a}cklund transformation for equation (\ref{sine1}) is
\begin{eqnarray}
\partial _{+}(\frac{\phi _{1}^{(0)}-\phi ^{(0)}}{2}) &=&\frac{m\lambda}{\beta} \sinh \beta(\frac{%
\phi _{1}^{(0)}+\phi ^{(0)}}{2}),  \label{PBT1} \\
\partial _{-}(\frac{\phi _{1}^{(0)}+\phi ^{(0)}}{2}) &=&\frac{m}{\beta\lambda} %
\sinh \beta(\frac{\phi _{1}^{(0)}-\phi ^{(0)}}{2}),  \nonumber
\end{eqnarray}
and first order correction to equation (\ref{PBT1}) is
\begin{eqnarray}
\partial _{+}(\frac{\phi _{1}^{(1)}-\phi ^{(1)}}{2}) &=&m\lambda(\frac{\phi
_{1}^{(1)}+\phi ^{(1)}}{2})\cosh \beta(\frac{\phi _{1}^{(0)}+\phi
^{(0)}}{2}),
\label{PBT2} \\
\partial _{-}(\frac{\phi _{1}^{(1)}+\phi ^{(1)}}{2}) &=&\frac{m}{\lambda }(%
\frac{\phi _{1}^{(1)}-\phi ^{(1)}}{2})\cosh \beta(\frac{\phi _{1}^{(0)}-\phi ^{(0)}%
}{2}).  \nonumber
\end{eqnarray}
The integrability condition of equations (\ref{PBT1}) and
(\ref{PBT1}) yields equations  (\ref{sine1}) and (\ref{sine2}),
respectively. To solve equation (\ref{sine1}) we first reduce the
problem to a one-dimensional problem by assuming that solution of
equations of motion are independent of time. The first soliton
solution of equation (\ref{sine1}) is
\begin{equation}
\varphi ^{(0)}=\frac{4}{\beta }\tanh^{-1}\exp (2mx),  \label{so1}
\end{equation}
the corresponding first order correction term to the solution is
\[
\varphi ^{(1)}=\frac{1}{\sinh(2mx)}.
\]
This solves the equation of motion and constraints for the
noncommutative sinh-Gordon equation to first order in $\theta$ .
\section{Conclusions}\label{con}

In summery, we have investigated a noncommutative version of
sinh-Gordon equation and discussed some of its properties as an
integrable equation. This noncommutative sinh-Gordon equation
reduces to an ordinary sinh-Gordon equation and constraints of the
model vanish in the commutative limit. The noncommutative version
of the linear system (or equivalently zero-curvature
representation) and Lax representation give an integrable
noncommutative sinh-Gordon equation. The constraints of the model
appear as total derivatives. We have also analyzed  the
integrability of the equation at perturbative level. We have
presented a set of B\"{a}cklund transformation for the zeroth
order sinh-Gordon equation and the first order correction to the
zeroth order B\"{a}cklund transformation. The soliton solution of
the equation has been obtained. We have also shown that the $1-$
soliton solution of the noncommutative sinh-Gordon equation solves
the equations of motion and its constraints.
\bigskip

{\large \bf{Acknowledgements:}} \\We acknowledge the enabling role
of the Higher Education Commission, Pakistan and appreciate its
financial support through ``Merit Scholarship Scheme for PhD
studies in Science \& Technology (200 Scholarships)''. We also
acknowledge CERN scientific information Service (publication
requests).

\end{document}